\documentclass[sigconf,9pt]{acmart}
\usepackage{svg}
\usepackage{algorithm2e}
\usepackage{algpseudocode}
\usepackage{xspace}

\AtBeginDocument{%
  \providecommand\BibTeX{{%
    \normalfont B\kern-0.5em{\scshape i\kern-0.25em b}\kern-0.8em\TeX}}}

\copyrightyear{2022}
\acmYear{2022}
\setcopyright{rightsretained} %

\usepackage{etoolbox}
\makeatletter
\patchcmd{\maketitle}{\@copyrightpermission}{
   \begin{minipage}{0.4\columnwidth}
     \href{http://creativecommons.org/licenses/by/4.0/}{\includegraphics[width=0.95\textwidth]{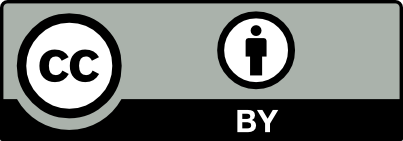}}
   \end{minipage}\hfill
   \begin{minipage}{0.6\columnwidth}
     \href{http://creativecommons.org/licenses/by/4.0/}{This work is licensed under a Creative Commons Attribution International 4.0 License.}
   \end{minipage}

   \vspace{5pt}
}{}{}

\makeatother

\acmConference[SIGMOD '22]{Proceedings of the 2022 International Conference on Management of Data}{June 12--17, 2022}{Philadelphia, PA, USA}
\acmBooktitle{Proceedings of the 2022 International Conference on Management of Data (SIGMOD '22), June 12--17, 2022, Philadelphia, PA, USA}
\acmDOI{10.1145/3514221.3520168}
\acmISBN{978-1-4503-9249-5/22/06}

\newcommand{\ex}{\textit{e.g.,}\ }
\newcommand{\ie}{\textit{i.e.,}\ }

\begin{document}
\fancyhead{}

\title{Demonstration of VegaPlus:\\Optimizing Declarative Visualization Languages}

\author{Junran Yang}
\affiliation{
  \institution{University of Washington}
  \city{Seattle}
  \state{WA}
  \country{USA}
}
\email{junran@cs.washington.edu}

\author{Hyekang Kevin Joo}
\affiliation{
  \institution{University of Maryland}
  \city{College Park}
  \state{MD}
  \country{USA}
}
\email{hkjoo@umd.edu}

\author{Sai S. Yerramreddy}
\affiliation{
 \institution{University of Maryland}
  \city{College Park}
  \state{MD}
  \country{USA}
}
\email{saiyr@umd.edu}

\author{Siyao Li}
\affiliation{
  \institution{University of Maryland}
  \city{College Park}
  \state{MD}
  \country{USA}
}
\email{siyaoli@umd.edu}
  
\author{Dominik Moritz}
\affiliation{
 \institution{Carnegie Mellon University}
  \city{Pittsburgh}
  \state{PA}
  \country{USA}
}
\email{domoritz@cmu.edu}

\author{Leilani Battle}
\affiliation{
  \institution{University of Washington}
  \city{Seattle}
  \state{WA}
  \country{USA}
}
\email{leibatt@cs.washington.edu}

\begin{abstract}
    While many visualization specification languages are user-friendly, they tend to have one critical drawback: they are designed for small data on the client-side and, as a result, perform poorly at scale. We propose a system that takes declarative visualization specifications as input and automatically optimizes the resulting visualization execution plans by offloading computational-intensive operations to a separate database management system (DBMS). Our demo emphasizes live programming of visualizations over big data, enabling users to write or import Vega specifications, view the optimized plans from our system, and even modify these plans and compare their performance via a dedicated performance dashboard.
\end{abstract}
\maketitle

\begin{CCSXML}
<ccs2012>
   <concept>
       <concept_id>10003120.10003145.10003151</concept_id>
       <concept_desc>Human-centered computing~Visualization systems and tools</concept_desc>
       <concept_significance>500</concept_significance>
       </concept>
   <concept>
       <concept_id>10010147.10010257</concept_id>
       <concept_desc>Computing methodologies~Machine learning</concept_desc>
       <concept_significance>300</concept_significance>
       </concept>
 </ccs2012>
\end{CCSXML}

\ccsdesc[500]{Human-centered computing~Visualization systems and tools}
\keywords{data visualization; scalability; dataflow}

\section{Introduction}

Developing interactive visualizations of large datasets requires significant effort. Besides making effective visualization design choices, the developer also needs to implement the underlying architecture to support interactivity at large scale, requiring expertise in client and server development, data management, and user interface design.
Ideally, visualization tools should be expressive enough to rapidly prototype a variety of designs, not only in terms of interface capabilities but also dataset scale and efficiency~\cite{moritz2015dynamic}.
Visualization specification languages such as D3~\cite{bostock_d_2011} and Vega~\cite{satyanarayan_reactive_2016} make the process of designing interactive visualizations more systematic, precise, and simple on the client-side. They often require the browser to load and process the data being visualized, which works well for designing responsive interfaces for small data. However, when handling data that is too large for the browser, these languages lack built-in support for coordination between client- and server-side data management and processing. 
Existing visualization systems like imMens, Falcon, and Kyrix address this problem for specific dataset and interface scenarios (e.g., geospatial datasets, crossfilter interfaces)~\cite{battle2020structured}, but they fail to support the level of expressiveness that D3 and Vega provide for innovative designs.

We propose an alternative solution that increases scalability while also aiming to preserve the flexibility of the underlying language.
Specifically, we present a series of visualization- and interaction-aware optimizations that can be integrated directly with existing visualization specification languages.
Our approach automatically determines which computations to keep on the client and which to offload to a separate DBMS, minimizing unnecessary network data transfers. Furthermore, our optimizer leverages knowledge of the compiled language structure to adapt and deploy database optimizations for interactive exploration contexts. In this way, we can reap the computational benefits of DBMSs while making it significantly easier for users to connect visualization tools with data processors that are already available to them.
We demonstrate our approach by implementing it for Vega; we call the resulting system \emph{VegaPlus}.
However, our approach can generalize to other declarative visualization languages as well.

Building a hybrid client-server application via a declarative language (Vega) is a promising approach for the following reasons. 
First, the underlying visualization system can automatically generate the necessary
client and server components. Vega's declarative design enables us to easily reason about and modify the data transformations that the Vega runtime executes. It decouples specification from runtime execution that utilizes a dataflow graph model, 
providing optimization opportunities via partitioning dataflow operators to transfer data and computation across client and server. As a result, the visualizations remain lightweight, stand-alone, and agnostic of the optimization work behind. Second, Vega is also expressive enough to capture the computational complexity of most visualization interfaces, including those tested in recent DBMS benchmarks designed for visual exploration scenarios~\cite{eichmann_idebench_2020,battle2020database}. Third, Vega is the backbone of a popular ecosystem of visualization tools, including Vega-Lite~\cite{satyanarayan_vega-lite_2017}, Voyager~\cite{2017-voyager2}, and Falcon~\cite{moritz_falcon_2019}; thus, making improvements to Vega is of interest to thousands of data enthusiasts, researchers, and companies worldwide. 

In this demo, users can use VegaPlus to implement scalable visualizations at no additional effort: they can focus on making nice visualizations, while VegaPlus automatically makes performance decisions for them. VegaPlus takes a dataset and a user-defined Vega specification as inputs and automatically loads the data into a user-selected DBMS. Its optimizer partitions the dataflow workload across client and server to minimize latency, both at initialization and upon user interactions. Furthermore, it comes with a dedicated performance dashboard that users can explore. The dashboard includes a graph visualization showing how the underlying execution plan is partitioned across client and server, as well as tooltips showing the details behind the nodes in execution plan. Users can also try their own ways to partition the execution plan by modifying the server/client partitioning scheme displayed in the dashboard, and compare it with our optimizer's performance. 

\begin{figure}
    \includegraphics[width=.75\linewidth]{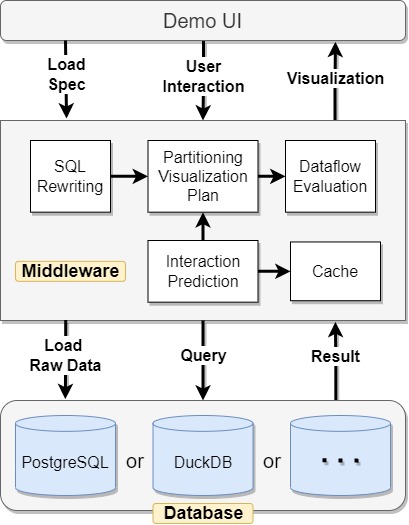}
     \caption{Demo architecture and workflow. %
     }
    \label{fig:demo}
    \vspace{-1em}

\end{figure}

\section{System Overview}
\label{sec:system-overview}

This section reviews the Vega dataflow and summarizes the system architecture (see \autoref{fig:demo}). The UI will allow users to create visualizations, visualize and interact with the visualization plans generated by the middleware, and compare the performance with their customized plans. Specifications and interactions are passed to our middleware, which (1) automatically instantiates a dataflow graph containing SQL translations for inner data operations from the declarative specification, (2) dynamically optimizes the partitioning of visualization plans, (3) prefetches data in anticipation of the following interactions and coordinates the cache, (4) evaluates the dataflow and handles communication across the client and server components. 
\vspace{-1em}

\subsection{Dataflow} \label{sec:dataflow}
The \textit{dataflow} is a common data model in visualization systems (\ex Vega~\cite{satyanarayan_reactive_2016}, VTK~\cite{schroeder_design_1996}) where its operators form a directed graph. A dataflow graph executes a series of data transformations (\ie processing a data stream) through its operators (\ex filter, map, aggregate) before the result is mapped to visual encodings. In Vega, a dataflow is automatically constructed based on the user's declarative specification. Streaming data objects pass through the edges and are processed by the operators. Parameters that define an operator can either be fixed values or live references to other operators. Interaction events update operator parameters or data inputs, and the changes are only re-evaluated by the necessary operators. 

\begin{figure*}
    \includegraphics[width=\textwidth]{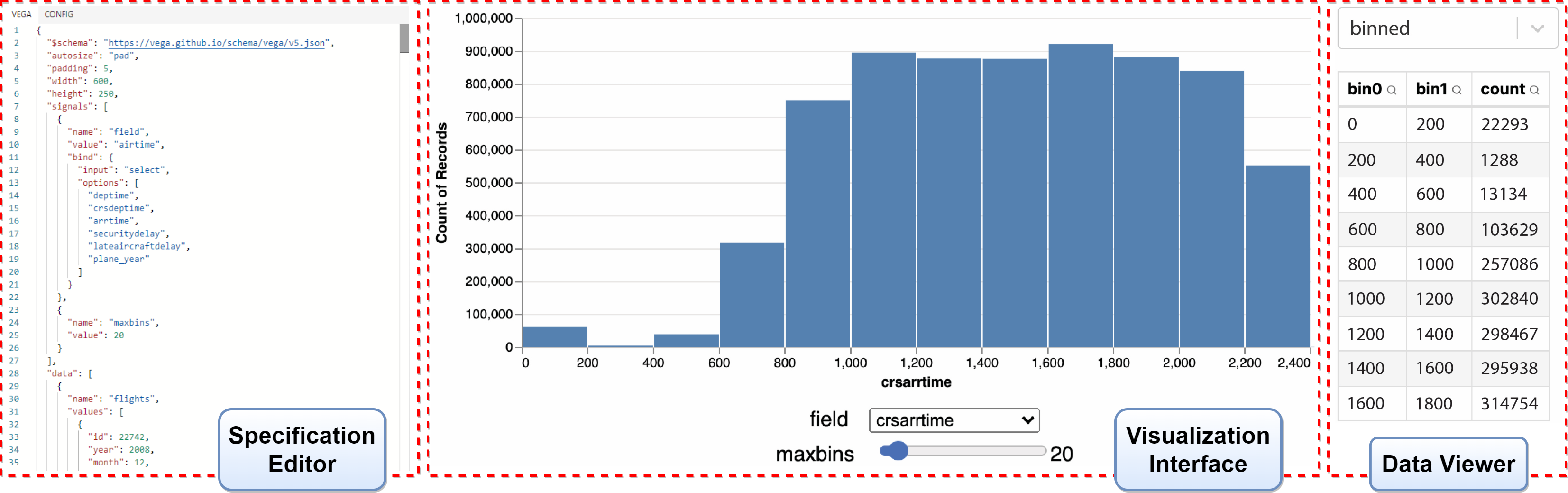}
    \caption{Visualization Editing View presenting the US Airline Flights example.}
    \label{fig:main panel}
\end{figure*}

\subsection{Middleware Optimization Dynamic}

Inspired by previous work in \cite{moritz2015dynamic}, VegaPlus contains a middleware server that takes the user-provided declarative visualization specification and instantiates an optimized dataflow graph in which operations are partitioned across client and server. VegaPlus optimizes how to partition the dataflow based on the dataflow graph, estimated data sizes, and current network latencies. The optimization dynamic is described in the following steps, where steps 1-3 handle startup (\ie visualization creation) and a loop is introduced in step 4 to handle changes and redundancies in the pipeline due to user interactions.

\textbf{(1) SQL rewriting:} A visualization specification can be roughly seen as processing raw data to the intended format so that the visual encoding specification maps processed attributes to target visual component properties. We assume rendering is not the dominant overhead due to the data reduction resulting from data processing and transformation, which in Vega is achieved via transform operators. To extend the dataflow model to a client-server architecture to use the scalability advantage of DBMSs, we automatically translate each transform operator to SQL queries and provide an extension to offload intensive calculations to the DBMS. The SQL translations are used in further optimization steps to decide whether they will be eventually executed in a DBMS, or their equivalent transform operations will be carried out by the client. 

\textbf{(2) Partitioning visualization plans:} For datasets of up to millions of records, client-side visualization tools are able to perform fast data processing and visualizing entirely on the client. For large datasets that cannot be fully loaded or fast-processed in the browser, raw data and computations can be offloaded to a backing DBMS. Based on a small experiment we conducted, for datasets with 4M rows Vega is faster than VegaPlus when it's not optimized, for 4M-10M performance is comparable and for 10M+ VegaPlus is much faster. 
With the raw data on the server-side, when to bring the dataflow back to the client-side remains an important problem for optimizing the overall cost and minimizing the latency. For static visualizations (or the initial visualization) with overly large data, the optimal point to split the dataflow is after the entire data processing to minimize the network cost, since the encoding mapping and rendering right after it don't dominate the latency. 

\textbf{(3) Optimizing server queries: } We optimize the part of the plan that is assigned to the server by node merging and SQL statement rewriting. By merging the SQL queries from individual operations, we can avoid unnecessary network round trips for data transfers. As for SQL statement rewriting, we optimize the subqueries by applying standard rule-based optimizations including pushing down derived conditions from outer subqueries, pruning projections, simplifying expressions, etc. 

\textbf{(4) Prefetching and re-partitioning:} Interactions impose an even stricter latency requirement for visualizations~\cite{battle2020database}. Based on the idea of partial execution in \autoref{sec:dataflow}, partially processed data can be brought back to the client earlier so that a downstream interaction parameterized by such data will only trigger a faster partial execution. To optimize partitioning for each interaction, we construct a prediction model with user modeling techniques~\cite{battle_dynamic_2016} for potential user actions, prefetch and cache requested data during idle times. Based on the prediction, we re-partition the dataflow to generate potential plans that split right before the interaction handlers in dataflow. When a interaction triggers, we pick the plan based on the interaction and cache state, and evaluate it accordingly.

\section{Demonstration Walkthrough}
In this section, we describe our demonstration scenarios. Our demo comes with several real-world datasets and commonly used visualization designs. In addition, the examples demonstrated are augmented by real ones from Vega visualization creators. Users will also be able to customize various types of visualizations using any dataset they choose. 

Our tool is motivated with the goal of enabling web-based interactive visualization of large data with minimum edits to the original Vega specification. For sake of space, we describe how to achieve this goal with just two of our various examples. 

 \textbf{US Airline Flights\footnote{https://www.transtats.bts.gov/OT\_Delay/OT\_DelayCause1.asp}:} The dataset consists of flight arrival and departure details for all commercial flights in the USA from 1987 to 2008. We use a simple record count histogram to explore the data distribution in terms of each data fields. Users can select the target data field from a drop down menu and use the slider to find a desired binning range to summarize the record count. 
 
 \textbf{Census-based Occupation History\footnote{https://www.census.gov/data/developers/data-sets.html}:} The dataset is comprised of the details collected by Census of the U.S. occupations reported between 1850 and 2000. In order to show all of the occupations reported in the year, aggregate transform operation was used by stacking each occupation atop each other, length of which was determined by its frequency in the respective year. The data are illustrated with an interactive stacked area graph, in which the user can perform multiple tasks. Filtering by gender can be performed by means of radio. Users can also use our search box that supports regular expressions to filter the jobs. 

\subsection{Demo Workflow}
\begin{figure*}
    \includegraphics[width=0.8\textwidth]{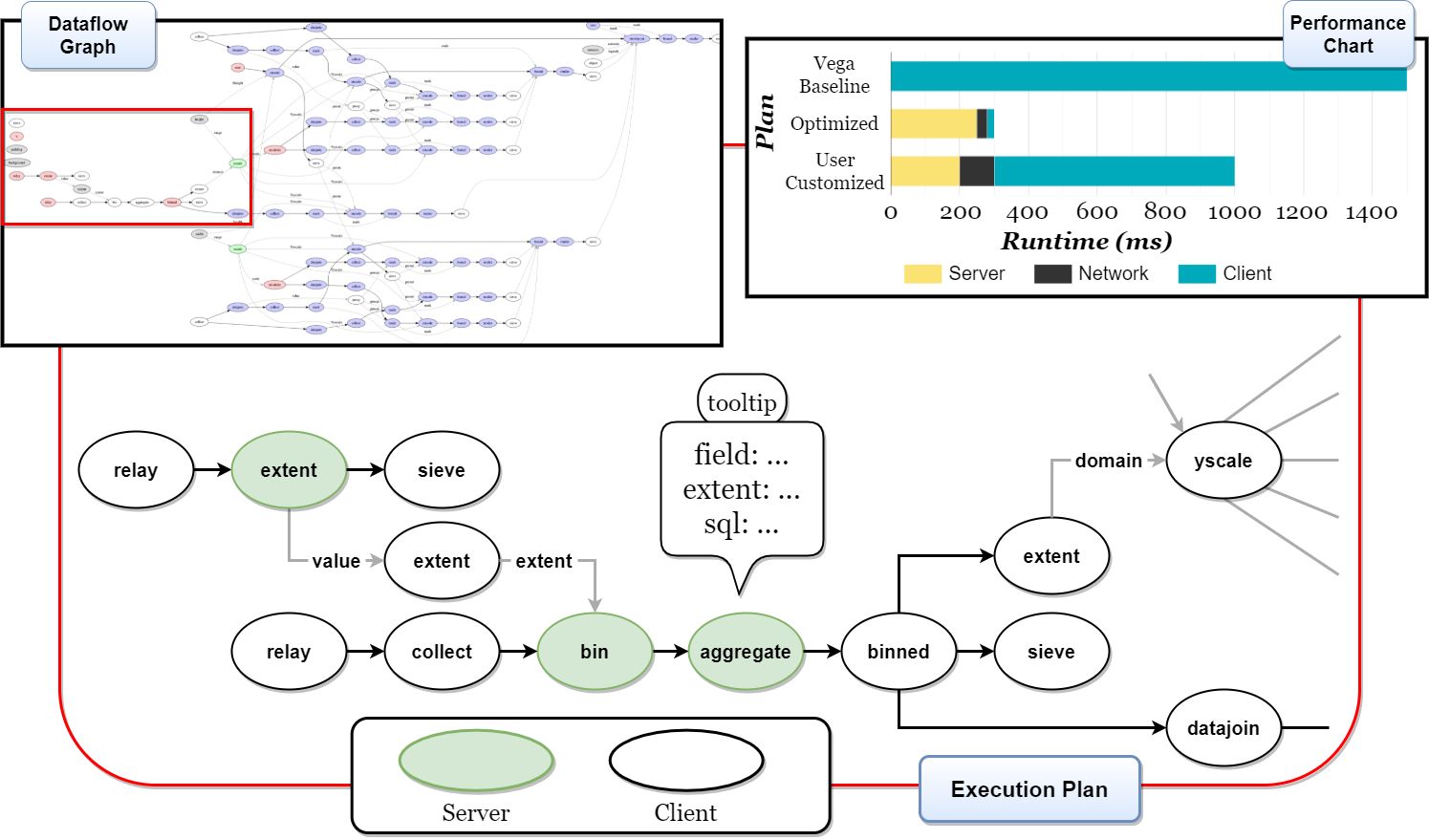}
    \caption{Performance View}\label{fig:dataflow}
\end{figure*}

Our demo interface consists of two major views: a \textit{visualization editing view }(\autoref{fig:main panel}) and a \textit{performance view }(\autoref{fig:dataflow}). In this section, we summarize how SIGMOD attendees can interact with each view and their significance in showcasing the contributions of our new visualization-aware optimizations.

\textbf{Visualization Editing View:} This view allows the user to create on-the-fly visualizations over large data, as well as inspect intermediate data results. It can be used as an independent visualization tool that enables fast visualization authoring and interaction with large data.
Users can create visualizations using one of our pre-loaded datasets and choose a DBMS back-end. We currently support PostgreSQL~\cite{postgres}, OmniSciDB, and DuckDB~\cite{raasveldt_duckdb_2019}.

To use the visualization editing view, the user uploads a specification and/or uses the live editor (\autoref{fig:main panel}-left) to modify the current specification. Existing example specifications (including the above examples) will also be made available to users. When modifying a specification in the editor, the user will see the updated visualizations rendered live (\autoref{fig:main panel}-middle). The flight example is shown in \autoref{fig:main panel}, where the user can also inspect the data results defined in the specification. For example, the ``binned`` data (\autoref{fig:main panel}-right) is resulted from the aggregation and is used to be mapped to the \textit{rect} visual marks (\ie bars).  %

\textbf{Performance View:} This view is an interactive dashboard of the overall visualization plan, which shows both the dataflow graph overview (top-left of \autoref{fig:dataflow}), execution plan and its performance chart (top-right of \autoref{fig:dataflow}). The main view at the bottom presents the dataflow graph; specifically, we show which operators are placed on the server and which are not by different colors. The execution plan in \autoref{fig:dataflow} continues the flights example, where the extent, bin, and aggregate 
operators are all placed on the server. Operator parameters and rewritten SQL queries will be shown as tooltips when users hover on the nodes. Further, users will be able to toggle on the operators to customize the partitioning.  For instance, the user could assign the bin operator to be executed on the client. In this case, data will be requested from the DBMS so that they can be allocated into buckets on the client, which will make the execution much slower because of more data transferring and inefficient SQL queries. Additionally, for users' reference, we show a stacked bar chart (top-right of \autoref{fig:dataflow}) to display the overall result and which components (\ie client, server or network) are taking up the most time. We will have one bar for each plan, and, for each stacked bar, we map different colors to the server, client, and network components. The user can compare the performance of Vega alone, our recommendation, and the user's own partitioning made by interacting with the dataflow graph or by simulating different network latencies. 
\vspace{-1em}
\section{Conclusion}
We present a new approach to scalable interactive visualization by optimizing declarative visualization specification languages. We demonstrate our approach by applying it to the Vega visualization language, and we refer to the resulting system as \emph{VegaPlus}.
Our proposed demonstration provides an intuitive, real-time experience with implementing and interacting with visualizations over large data, and a unique environment for interactively optimizing the underlying execution plans that is easily adjustable for demo users. 

\section{ACKNOWLEDGMENTS}
The authors wish to thank the UWDB group, the BAD Lab, the IDL, and our paper
reviewers for their thoughtful feedback. This work was supported in
part by NSF award IIS-1850115. 

\bibliographystyle{ACM-Reference-Format}
\bibliography{sample-base}  %
\end{document}